\documentclass[11pt,preprint,flushrt]{aastex631}

\usepackage{natbib}
\usepackage{graphicx}
\usepackage{amsmath}
\usepackage{multirow}

\received{January 2023}
\revised{May 2023} 
\accepted{May 2023}

\shorttitle{Asteroid Albedos}
\shortauthors{Murray}
\graphicspath{{./}{figures/}}

\begin{document}

\title{Using neural networks to model Main Belt Asteroid albedos as a function of their proper orbital elements} 

\correspondingauthor{Zach~Murray}
\email{zachary.murray@cfa.harvard.edu}
\author[0000-0002-8076-3854]{Zachary Murray}
\affil{Center for Astrophysics | Harvard \& Smithsonian, 60 Garden Street, Cambridge, MA 02138}

\begin{abstract}
    Asteroid diameters are traditionally difficult to estimate.  When a direct measurement of the diameter cannot be made through either occultation or direct radar observation, the most common method is to approximate the diameter from infrared observations. Once the diameter is known, a comparison with visible light observations can be used to find the visible geometric albedo of the body. One of the largest datasets of asteroid albedos comes from the NEOWISE mission, which measured asteroid albedos both in the visible and infrared. We model these albedos as a function of proper elements available from the Asteroid Families Portal using an ensemble of neural networks.  We find that both the visible and infrared geometric albedos are significantly correlated with asteroid position in the belt and occur in both asteroid families and in the background belt. We find that the ensemble's prediction reduces the average error in albedo by about 37\% compared to a model that simply adopts an average albedo, with no regard for the dynamical state of the body. We then use this model to predict albedos for the half million main belt asteroids with proper elements available in the Asteroid Families Portal and provide the results in a catalog.  Finally, we show that several presently categorized asteroid families exist within much larger groups of asteroids of similar albedos - this may suggest that further improvements in family identification can be made. 
\end{abstract}



\section{Introduction}

One of the most fundamental properties of an asteroid is its diameter; it is a prerequisite to compute an asteroid's bulk density if the asteroid's mass is known, or conversely, can be used to infer a mass - if its density can be otherwise estimated.  However, the diameter of an asteroid is often difficult to estimate.  Most surveys of asteroids occur in the visible portion of the spectrum. These surveys record a flux from the object which - when paired with knowledge of its orbit - can be used to derive an absolute or H magnitude.  Unfortunately, since the observed flux of the asteroid in these wavelengths is reflected light, an estimate of the diameter of an object cannot be made without knowledge of the asteroid's albedo.  

Given this limitation, several different methods have been developed to more accurately constrain asteroid albedos and diameters. One of the most direct methods is through occultations~\citep[e.g.,][]{Tanga_2007,Herald_2020}.  In such an event, an asteroid passes in front of a distant star, blocking its light from the perspective of an observer for a few seconds. The asteroid's diameter can then be directly estimated from the duration of occultation and knowledge of the asteroid's orbit. Unfortunately, occultation events are relatively rare, since they require precise alignment of the asteroid and background star. Hence, occultation-derived diameters are not available for most asteroids. 

A second way to directly measure an asteroid's diameter is through radar observations. Such observations are often of sufficiently high resolution to detect the angular extent of an asteroid outright, which makes calculating the diameter straightforward~\citep[e.g.,][]{Ostro_2000,Ostro_2006}. However, the inverse square law limits radar observations to only relatively close asteroids, consequently the population of asteroids that can be directly measured in this way is also rather small. 

Finally, there are infrared observations.  At infrared wavelengths the asteroid's flux is primarily thermal emission and is proportional to its surface area \citep[e.g.,][]{Allen_1971}.  Hence, an estimate of the diameter can be derived from these observations, with the albedo being estimated by comparison with the visible reflected flux.  In practice, this method is by far the most productive way of estimating diameters, as infrared surveys can easily cover large portions of the sky. The largest of these surveys is the WISE/NEOWISE survey which has provided nearly 200,000 albedo measurements of solar system bodies, including $125,217$ unique main belt asteroids~\citep[e.g.,][among others]{Masiero_2011,Grav_2012,Masiero_2014}.  Other infrared missions have also contributed to the total of known albedos, including the IRAS mission \citep{Tedesco_2002}, AKARI survey \citep{Usui_2011} and the Spitzer Space Telescope~\citep{Gustafsson_2019}. This total, however, is a fraction of all the asteroids cataloged by the Minor Planet Center - which as of June 2022 number over a million.  Consequently, the albedos of most asteroids remain unknown.  This fraction will only decrease as new surveys discover more asteroids, with the Rubin Observatory alone projected to discover up to five million in the main belt~\citep{LSSTScienceCollaboration_2009}.

Motivated by the observation of \citet{Masiero_2014} that asteroids belonging to the same family tend to have similar albedos, we wish to understand how much information an asteroid's proper orbital elements contain about its albedo. We investigate this question by constructing a model to predict albedo as a function of proper orbital elements for asteroids in the main belt.  

Proper elements are essentially time-averaged Keplerian orbital elements that serve as quasi-invariants of motion.  They remain unchanged over long time scales~\citep{Lemaitre_1993}.  It is useful to contrast these with the osculating orbital elements, which are the instantaneous Kepler elements of an asteroid at a given time.  These elements change over timescales as short as a few thousand years (except for the mean anomaly which changes over a single orbit), with the change primarily being driven by gravitational perturbations from other planets and asteroids.   Although the instantaneous difference between the two elements is typically small \citep[see examples in][]{Knezevic_2002}, the stability of the proper orbital elements make them particularly suitable for studying asteroid families~\citep[e.g.][]{Hirayama_1922,Lindblad_1971,Zappala_1990,Zappala_1995}.  These families are often tightly clustered in the space of proper elements, whereas examining the same family in the space of osculating elements results in much looser clustering, due to the short timescale differential change in the elements. 

A variety of different methods to compute proper orbital elements have been developed, including analytic models \citep{Yuasa_1973,Knezevic_1989,Milani_1994}, semi-analytic models \citep{Lemaitre_1994,Gronchi_2001,Fenucci_2022} and numerical approaches \citep{Knezevic_2000}. In this paper, we concern ourselves with the proper elements provided by the Asteroid Families portal site \citep{Novakovic_2019}.

\section{The Model}
\label{sec:model}
The Asteroid Families Portal provides proper elements for nearly $600,000$ numbered main belt asteroids, computed with the methods of \citep{Knezevic_2000,Knezevic_2003}, and family classifications derived using the hierarchical clustering method developed in \citep{Radovic_2017}. In our sample, we exclude the active main belt objects, as their activity may result in variations in the measurement of their fluxes or cause their albedos to change over time. We use the results of the NEOWISE asteroid survey as our training data set, as it is currently the largest such set of data available~\citep{Mainzer_Bauer_Cutri_Grav_Kramer_Masiero_Sonnett_Wright_2019}.  The overlap between the two datasets is significant, not only including many main belt asteroids, but also the Hungaria and Hilda asteroids.   Notably absent here are the Jupiter Trojans, as proper elements for the Jupiter Trojans must be computed with different methods than those outlined above - with interaction with Jupiter taken into account.  In addition, \citep{Grav_2011} suggests these Trojans are largely homogeneous in albedo compared to the main belt.  Hence there is likely less insight to be gained by constructing a model that incorporates their dynamical states.  Our training sample thus consists of a total of 122,309 asteroids both have measured NEOWISE albedos and computed proper elements. Among these, several asteroids were observed multiple times, and their albedos were taken to be the average of the estimated albedos, with the corresponding uncertainty computed by quadrature-addition of the individual errors. There was also a small sample of asteroids with negative reported albedos - these were removed from the sample.

\begin{figure}[!ht]
\centering
    \includegraphics[totalheight=7.5cm]{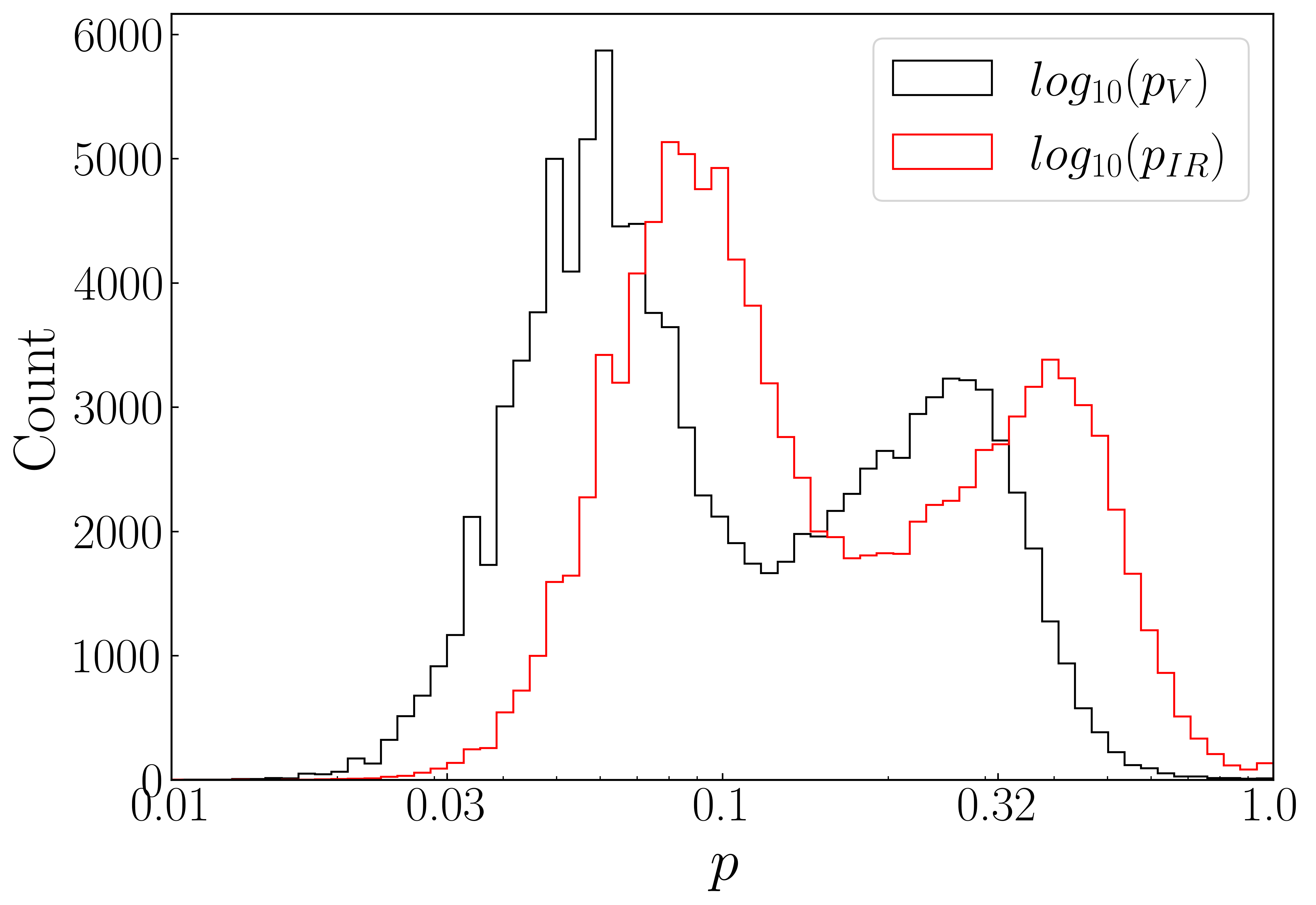}
    \caption{
    Here we show the distribution of the logarithm of the visible and infrared albedos from the NEOWISE catalog.  The distribution of both distributions are bimodal, with the visible albedos peaking near $p_V = 0.05$ and $p_V = 0.3$ and the infrared albedos peaking near $p_{IR} = 0.1$ and $p_{IR}=$0.4}
    \label{albedohist}
\end{figure}

As is common in the literature we work with the logarithm of the albedos rather than the albedos themselves.  This parameterization allows for more contrast between dark asteroids with similar albedos.  It is also particularly convenient for determining the diameter, as the logarithm of the visible albedo can be related directly to the logarithm of the diameter by
\begin{equation}
    \log_{10}(D) = 3.1236 - 0.5 \log_{10}(p_V) - 0.2 H,
\end{equation}
where $p_V$ is the visible geometric albedo, $H$ the absolute visible magnitude and $D$ the diameter in kilometers \citep[e.g][]{Harris_1997}. Fig (\ref{albedohist}) shows the distribution of the logarithms of the visible and infrared albedos of the main belt asteroids as measured by NEOWISE.  As noted by many previous studies \citep[e.g.][]{Morrison_1977,Masiero_2011,Usui_2013} the distribution of visual albedo in the main belt is bimodal, with peaks near $0.05$ and $0.3$, with the highest peak largely being from the contribution of darker asteroids in the outer main belt. The visible albedo $p_V$ and the infrared albedo $p_{IR}$ observed by NEOWISE are generally highly correlated with each other, with the former being smaller than the latter for most asteroids, similar to what has been found among the Trojan asteroids \citep{Grav_2011}. 

We model the variation in both albedos as a function of the proper eccentricity $e$, the sin of the proper inclination $\sin(i)$, and the proper semi major axis $a$.  We conducted a variety of preliminary tests by employing different machine learning techniques on the data set including, polynomial regression, the decision tree and random forest regressor algorithms in the \texttt{SCIKIT-Learn} python module \citep{sklearn_api}, in addition to neural nets implemented with TensorFlow \citep{tensorflow2015-whitepaper} to assess their suitability for making predictions on the data set.  We found that the decision tree and random forest algorithms performed worse than the neural nets as measured by the root mean squared error residual.  This is likely because predictions made by these algorithms are based off a discrete criterion, which is poorly suited for continuous variables like proper elements.  Polynomial regression, while continuous, lacked the needed complexity to fit the subtle detail in the belt.  Consequently, we use neural nets as our algorithm of choice for this data set. This approach has also shown promise in application to several other problems relevant to the study of small solar system bodies, including taxonomy \citep{Penttila_2021}, image classification \citep{Duev_2021} and dynamics \citep{Carruba_2022b,Carruba_2022c}.  We use a relatively shallow network consisting of six layers of nodes with the ``relu'' activation function implemented by TensorFlow.  We assess the quality of the fit with a root-mean-squared-error loss function and optimize using stochastic gradient descent. 

\begin{figure}[!ht]
\centering
    \includegraphics[totalheight=7.5cm]{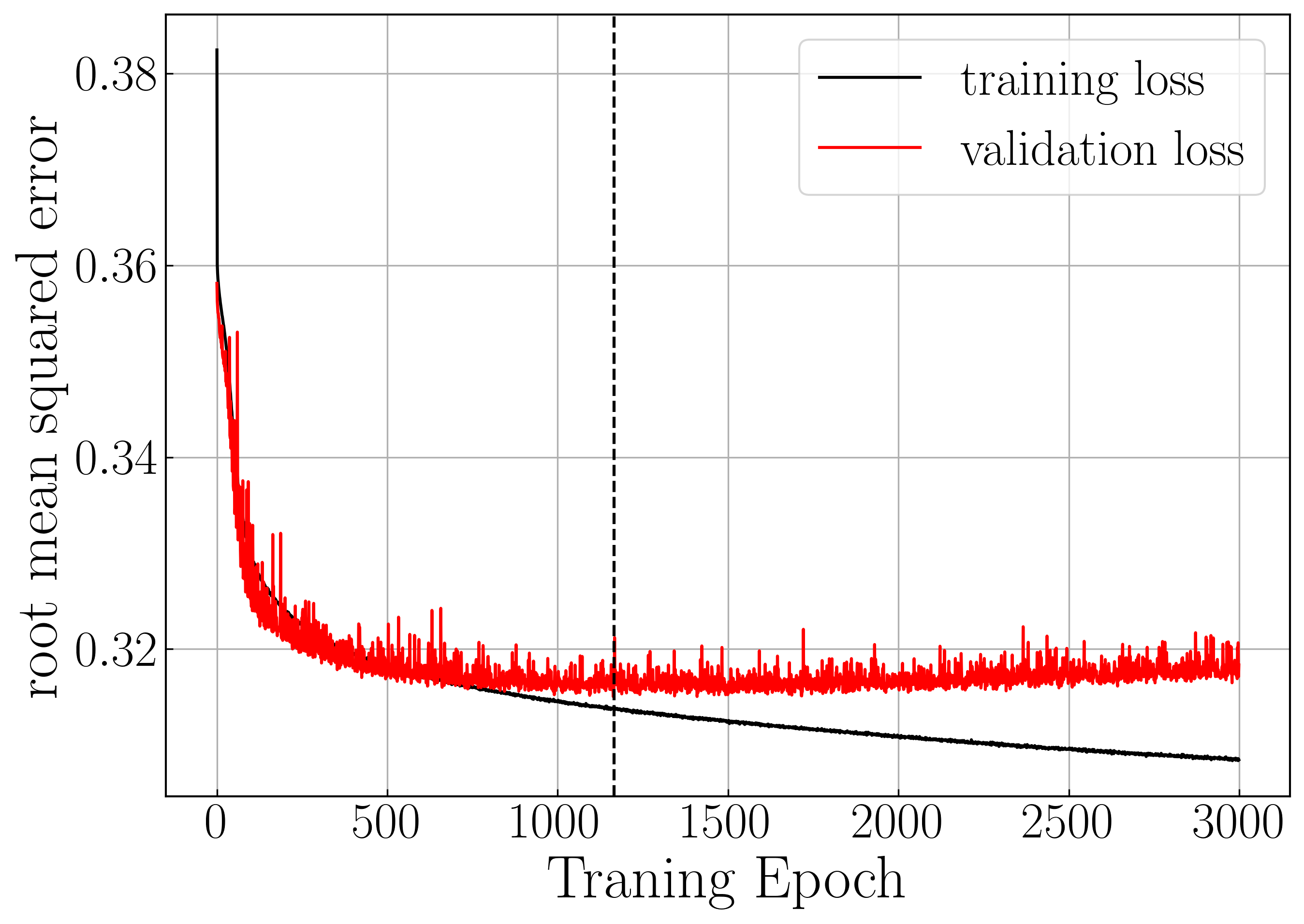}
    \caption{
    Here we show both the training (black) and validation (red) root mean square error vs epoch for a single representative sample of our ensemble of fits to the visible albedo.  The validation loss rapidly decreases at early epochs before starting to decrease more slowly after epoch $500$.  It reaches a shallow minimum near epoch $1100$ before gradually increasing and overfitting at later epochs. The dashed vertical line denotes the epoch of minimum validation loss. The weights from this epoch are used in the ensemble.}
    \label{testtrain}
\end{figure}

We simulate the effect of observation error by re-running the fitting procedure 1000 times on randomly generated samples of the data with each albedo being randomly chosen from a distribution with probability given by
\begin{equation}
     \begin{cases} 
          0 & x < 0 \\
         \frac{C}{\sigma\sqrt{2 \pi}} \exp({-\frac{1}{2}\frac{\left(x-\mu\right)^2}{\sigma^2}}) \ & 1 \geq x \geq 0 \\
          0 & x > 1,\\
      \end{cases}
\label{truncnormal}
\end{equation}
where $\mu$ is the albedo and $\sigma$ its corresponding error in the NEOWISE dataset. Here $C/2 = erf(\frac{1-\mu}{\sqrt{2}\sigma}) + erf(\frac{\mu}{\sqrt{2}\sigma})$ is a normalization constant, which can be written in terms of the error function ($erf(x) = \frac{2}{\sqrt{\pi}} \int_0^{x} e^{-t^2} dt$.  Equation \eqref{truncnormal} is a truncated normal distribution, which prevents unphysical albedos from being generated.  
We use a training-validation split fraction of $0.2$ for each fit and track both the loss on the training set and the validation set as a function of the training epoch.  As training proceeds, the loss on the test set decreases, while the loss of the validation set reaches a minimum after some time, as shown in Fig (\ref{testtrain}). After this point, the fit to the training set improves while the loss on the validation set beings to worsen, implying that the model is beginning to over-fit. To avoid over-fitting we terminate training our model when the error on the validation set is minimized and use the weights at this epoch for our predictions. We perform this procedure on both the visual albedos and the infrared albedos, since the former is useful for predicting the diameter of asteroids and the latter has shown promise for use as a proxy of asteroid taxonomy \citep{Masiero_2014}.

\section{Results}

\begin{figure}[p]
    \makebox[\linewidth]{
        \includegraphics[width=0.80\linewidth]{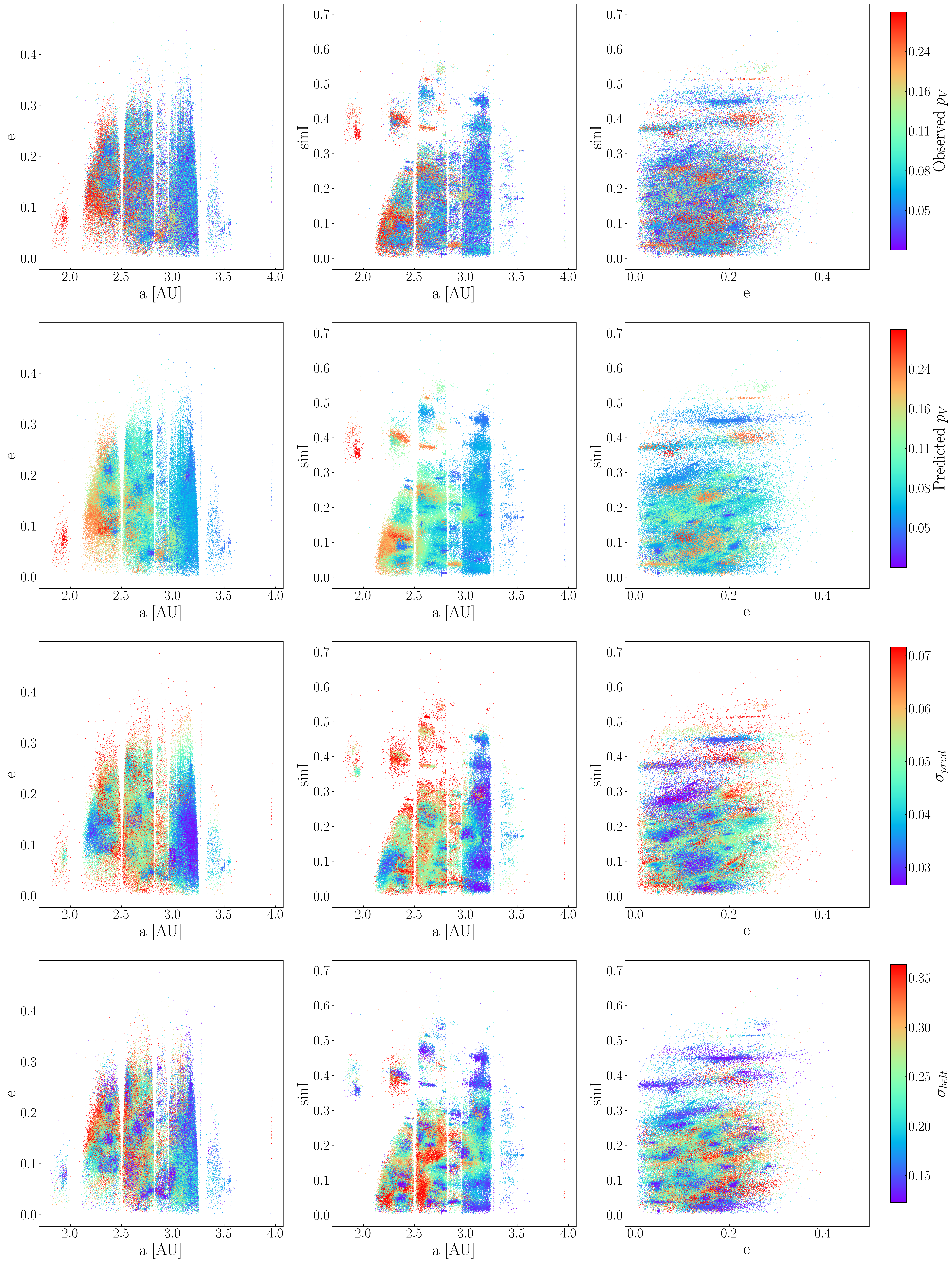}
    }
    \caption{Here we show the true visible albedos as reported by NEOWISE (top row), the mean predicted visible albedos (second row) and the averaged model uncertainty (third row) and the stochasticity of the belt (final row)  as a function of the proper elements (note the difference in color scales between each row). We can see large groups of asteroids, corresponding to different families, tend to have low albedos, these families make up a significant portion of the inner asteroid belt.  In addition, there is a clear, but highly structured gradient in albedo from the inner to outer main belt.  The smoothing in the predicted albedos makes some structure in the main belt more obvious, such as the large region inclined asteroids of asteroids with low albedo in the central main belt.   The third row shows that model uncertainties due to observational error tend to be largest in sparsely populated parts of the belt, whereas the final row shows that areas near asteroid families tend to be relatively homogeneous in albedo}
    \label{mainresult}
\end{figure}

We show a compilation of our results for visible albedo in Fig (\ref{mainresult}), including predictions and relevant errors.  While we will focus our discussion on the visible albedo $p_v$, the tight correlation between visible and infrared albedo ensure that all of the conclusions in this section generalize to the infrared albedo as well.  We generate our predictions by taking the average of the predictions made by our ensemble of neural nets on the training set. These predictions are shown in the second row of Fig (\ref{mainresult}) This prediction necessarily smooths the true albedos provided by NEOWISE.  As a result of the smoothing, significant structure in the main belt becomes more obvious, with large groups of asteroids with similar albedos clearly visible.  In addition, several other subtle trends, including an increase in average albedo with inclination in the middle belt, become more clearly visible.  Our averaging procedure also allows us to quantify the effects of measurement errors on our mean predictions.  Since each individual net in the ensemble fits a slightly different data set with simulated noise, the standard deviation of the individual predictions gives a numerical measure of the effect of NEOWISE measurement errors on the predictions of the ensemble.  As shown in the third row of Fig (\ref{mainresult}), measurement error tends to have the largest impact on the predictions in regions of the belt where there are few measured asteroids.  In dense regions, like the outer belt, there are enough observations that the prediction effectively averages the errors in the measured albedos of individual asteroids, leading to a small spread between models and higher fidelity predictions.  Overall, we find that the error in model prediction is relatively small, with $\sigma_{pred} < 0.05$ for most of the belt. 

While the spread in models gives us a notion of the sensitivity to measurement error, it does not take into account the error due to the inherent stochasticity of albedo in the belt. To quantify this we consider the absolute residuals between our averaged prediction and the NEOWISE measured albedo, and fit a neural network to it using the same parameters as explained in Section \ref{sec:model}, except for using a slightly higher test-train fraction of $0.3$ chosen since the residuals are noisier than the albedos themselves.  As can be seen in Fig (\ref{lresiduals}), these residuals are nearly symmetric and almost normally distributed, hence the distribution of the absolute residuals is very nearly that of a half-normal distribution.   Fitting with a neural net with our chosen loss function will act like an adaptive moving average, however, statistical distributions are typically described using their standard deviations. Hence, to convert the uncertainty due to stochasticity in the belt into $\sigma_{belt}$, we inflate the neural network predictions by a factor of $\sqrt{\pi/2}$, which is the appropriate factor for a half-normal distribution. The scaled predictions from this net are shown in the final row of Fig (\ref{mainresult}). $\sigma_{belt}$ is therefore the uncertainty in the neural net prediction of the logarithm of the albedo due to stochasticity in the belt.  This reveals that the parts of the belt near asteroid families and the outer belt tend to be homogeneous in albedo, further they reveal that the middle belt ($2.5-3.0$ AU) exhibits the largest diversity in asteroid albedo particularly at very low and very high $\sin(i)$.  These regions likely exhibit greater diversity as they correspond to areas where families of high albedo asteroids mix with those of lower albedo, (a correspondence that can be seen in Fig (\ref{mainresult}) this diversity results in a broader local distribution of albedo and a higher inferred $\sigma_{belt}$. Finally, we find the intrinsic stochasticity in the belt is larger, by nearly a factor of $5$, than the uncertainty in the mean prediction due to measurement error. Therefore, for most asteroids, the total error in the albedo prediction for a certain set of proper elements can be approximated by $\sigma_{belt}$ alone.  This implies that our results are largely insensitive to the measurement error and that future improvements in measurement accuracy will not have a large effect on the belt-wide predictions. 

Outside of the illuminating structure in the belt, the main use case for this model is to predict albedos of asteroids for which infrared or other observations aren't available but whose orbits are known.  These asteroids will be disproportionately small and dim.  Small bodies are subject to orbital perturbations from the YORP-Yarkovsky effects and radiation pressure, this can cause significant changes in their orbital semi-major axes and their proper elements.  These perturbations cause asteroid families to spread and hence cause small bodies to drift far from other bodies of similar albedos, weakening the correlations with the proper elements \citep{Bottke_2001}.

\begin{figure}[!ht]
\centering
    \includegraphics[totalheight=7.5cm]{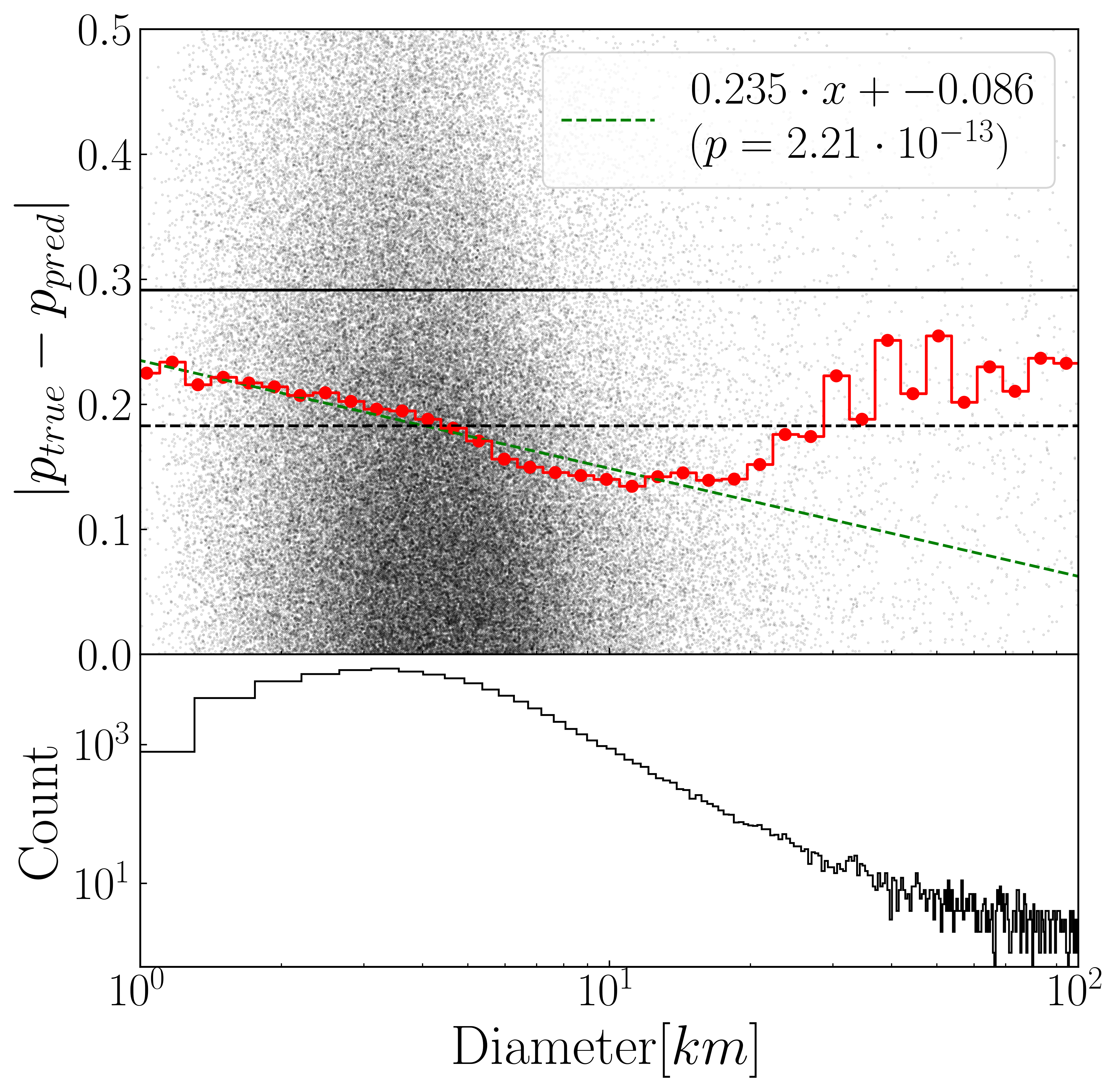}
    \caption{We illustrate the increase of mean absolute error among asteroids with small diameters. The top panel shows the absolute error vs the $\log_{10}(D)$ for each of the asteroids in our dataset (grey points) and a binned average of that error (red step function).  The behavior over $D < 20 \mathrm{km}$ is modeled by the linear function shown here as a dashed green line.  The horizontal dashed line is the mean absolute error over the entire dataset ($0.182$), whereas the horizontal solid line is the mean absolute error if the predictions were taken to be a simple mean of the asteroid albedos (without accounting for their orbital elements) ( $0.291$).  The neural network decreases the mean absolute error by just over $37$ \%.  The bottom panel shows a histogram of the grey points, there are very few asteroids in the sample with diameters of more than $20 \mathrm{km}$, and we ignore such bodies in our linear fit.}
    \label{diaerror}
\end{figure}

These effects suggest that the accuracy of our predictions may shrink for smaller asteroids, which are more susceptible to these, and other, effects.  To test for this we compute the mean absolute residual between our predictions and the NEOWISE observations as a function of the diameter. These diameters were included in the set of NEOWISE observations and were computed using the observed visible geometric albedos. We find the residual decreases approximately linearly with the $\log_{10}(D)$ for increasing D where $D$ is the asteroid diameter over the interval $D < 20.0 \mathrm{km}$, with the effect over this range being approximately described as $ |p_{true} - p_{pred}| = c_1 + \log_{10}(D) c_2$, with $c_1 = 0.2506$ and $c_2 = -0.0993$.  Since the residuals vary widely with diameter we also compute the $p$ value of the fit, assuming the null hypothesis that the slope is zero.  We find $p = 2.21 \cdot 10^{-13}$, hence our fit is significant.   This effect is illustrated graphically in Fig (\ref{diaerror}).  Outside these bounds, the mean absolute residual varies in a more stochastic fashion due to decreasing sample size. While we do observe that the model becomes less effective at predicting albedo for objects of small diameters, even our smallest asteroids show improvement compared to an approach where the asteroid's dynamical state is not taken into account. This implies that albedos of most small main belt asteroids are likely still significantly correlated with their proper elements.  This result increases our confidence that the model can generalize to even very small asteroids ($\approx 1 \mathrm{km}$) across the main belt. 

\subsection{The Role of Asteroid Families}

\begin{figure}[!ht]
\centering
    \includegraphics[totalheight=5.5cm]{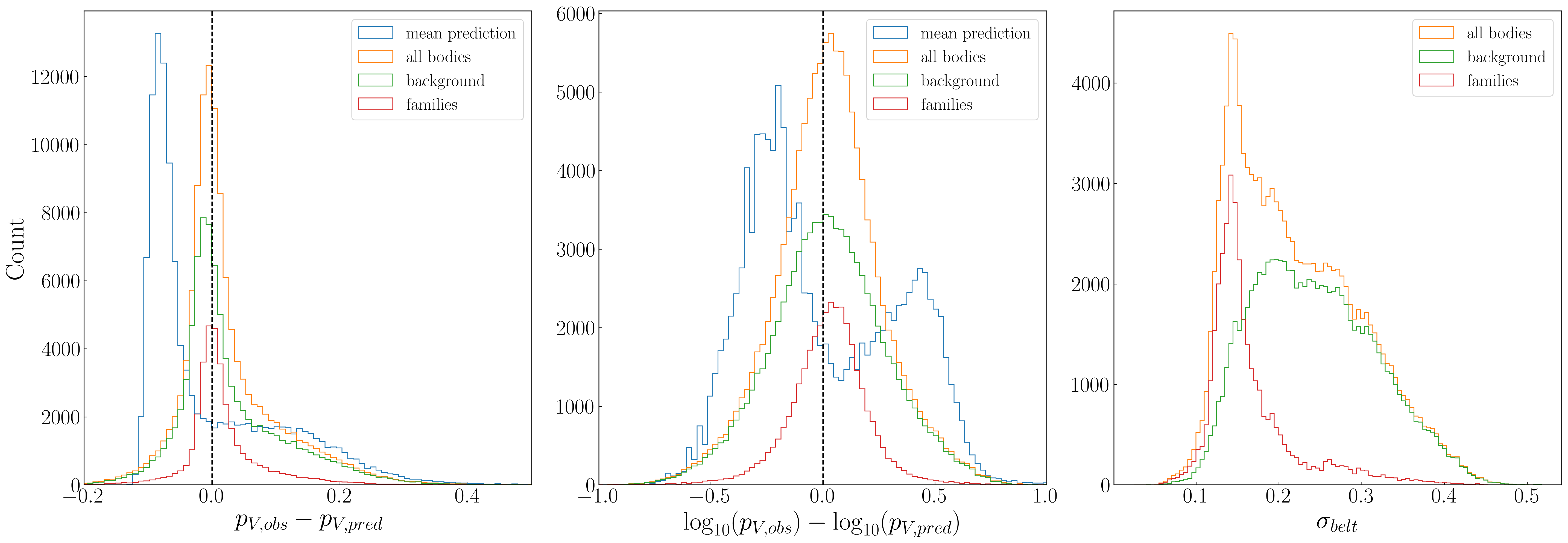}
    \caption{
    In the left panel, we show the residuals for the neural net for all the asteroids in our sample (all bodies), the residuals for those in recognized asteroid families (families) and those not known to belong to an asteroid family (background) and the residuals for a prediction based on the average albedo. We contrast them with a naive prediction that simply assigns the average albedo to every asteroid regardless of its dynamical state.
    We can see that the residuals from the neural net ensemble are significantly smaller than that of the naive prediction. In the right panel, we show the distribution of $\sigma_{belt}$ for asteroids in and outside families.  We recover the relative homogeneity of albedo in asteroid families compared to that of the background belt.   
    Note the small displacement of the residuals from zero comes from the errors being given in terms of the albedo itself, rather than its log.  The transformation between these causes a small asymmetry in the error which shows up here as a small shift away from zero.}
    \label{lresiduals}
\end{figure}

Given the homogeneity in albedo among asteroid families, it becomes natural to ask how much of our predictive power comes from correctly modeling the albedos of different families versus fitting subtle trends that extend across the belt. To test this we divide the asteroids into families using the Asteroid Families Portal classification scheme \citep{Milani_2014,Knezevic_2014}, and compute residuals independently for each. The results of this test are displayed in Fig (\ref{lresiduals}). These residuals show that asteroid families do have significantly smaller residuals than background asteroids.  However, there is still significant improvement in the residuals of background asteroids when compared to a naive prediction that assigns all asteroids the average albedo of the dataset. These background asteroids have highly inhomogeneous albedos relative to those in families. Hence, the efficacy of the neural net ensemble isn't solely due to correctly predicting the albedos of homogeneous families, but that it also effectively fits subtle trends in the highly inhomogeneous background asteroids of the main belt. 

\begin{figure}[!ht]
\centering
    \includegraphics[totalheight=7.5cm]{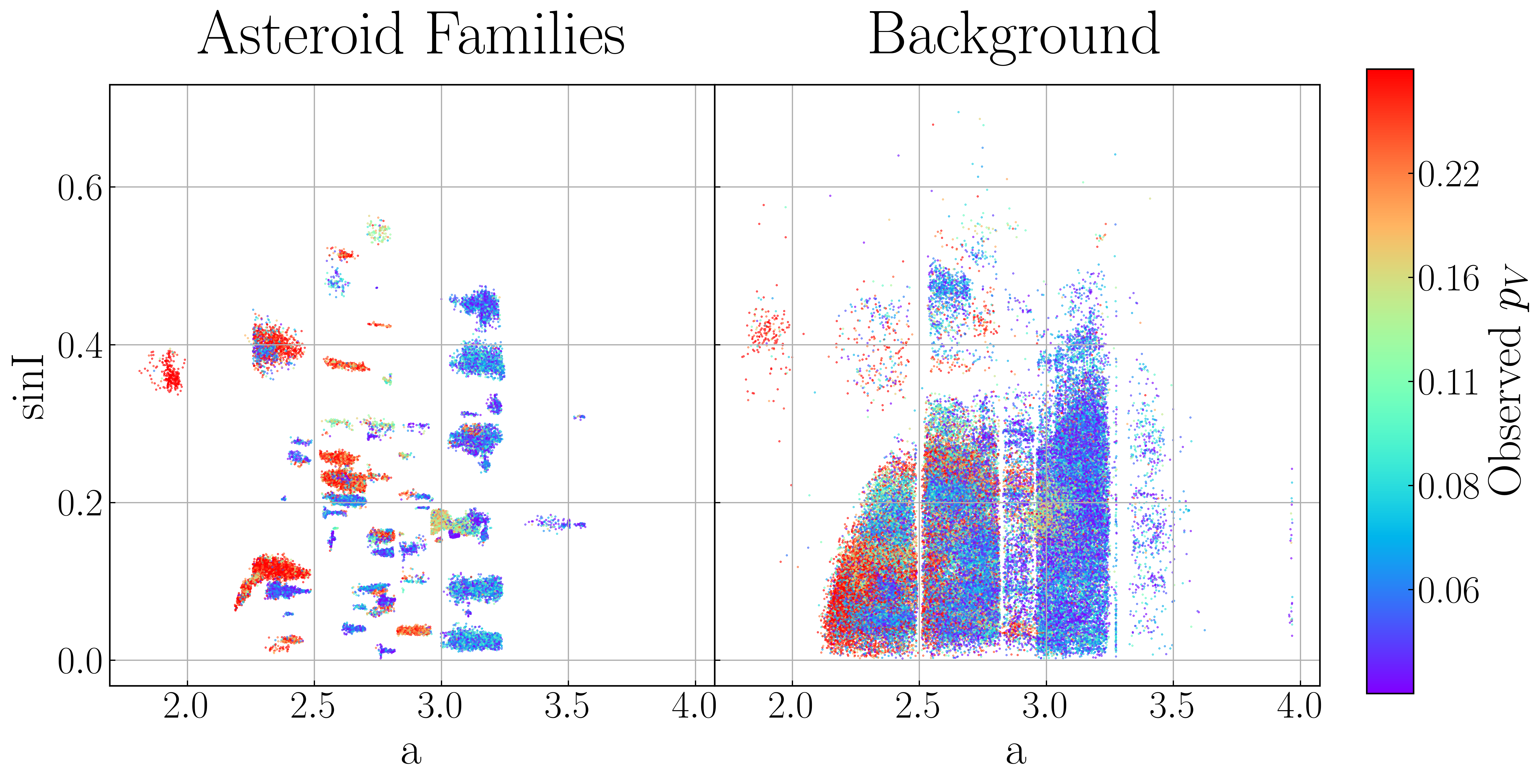}
    \caption{
    We show $\log_{10}(p_V)$ asteroid families in the left panel, and the background of asteroids not in families in the right panel, shown in the plane of $\sin(i)$ and $a$ proper orbital elements.  We can see the classified families only comprise a portion of the regions with similar albedos. }
    \label{familyvis}
\end{figure}

Finally, in addition to being generalizable, our model smooths the underlying structure observed by NEOWISE and reveals subtle structure in the belt. For example, as shown in Figure (\ref{familyvis}), homogeneity in albedo is not limited strictly to asteroid families.  Asteroids located near identified families - but not in them - often share similar albedos with asteroids in those families, behavior would be expected if these families form by collisional disruption of large asteroids \citep{MARZARI_1995,Bottke_2005,Milani_2014,Spoto_2015}.  For example, the Erigone family (near $a=2.4$, $\sin(I)=0.09$ in Figure \ref{familyvis}) exists within a much larger dynamical grouping of low albedo asteroids. While this tendency does not affect the quality of our fits, as they are agnostic to the family classification, it does beg for explanation.  It seems likely that the Asteroid Families Portal family classification is incomplete, or that certain collisional families overlap with larger dynamical families of similar properties. This might imply that larger numbers of asteroids are members of families than currently thought.  This is consistent with previous work from \citep{Broz_2013,Carruba_2013,Carruba_2013b,Carruba_2022} which find `halos' of asteroids with similar properties but are often not found by traditional hierarchical clustering methods.

It's also worth examining how our results extend to features of the belt we have not trained on. One such test is to examine or predictions over the $z_2$ resonance near the Erigone family. The $z_2$ resonance is defined by a commensurability of the precession frequencies of the asteroids contained in it and Saturn.  In this case $2 \dot \omega + \dot \Omega = 2 \dot \omega_6 + \dot \Omega_6$  where $\dot \omega$ and $\dot \Omega$ are the precessional frequencies of the asteroids and $\dot \omega_6$ and $\dot \Omega_6$ are those of Saturn, in this case taken from \citep{Carruba_2009}.  We pick out asteroids near the $z_2$ secular resonance by considering only those asteroids with semi-major axes between $2.3$ AU and $2.45$ AU, $e \approx 0.2$ and $\sin(i) \approx 0.1$. When points are plotted in the space of their semi-major axis vs relevant precessional frequency, the secular resonance corresponding to the Erigone family appears as a horizontal line and can be see in Figure \ref{familyvis}.  We can now compare our predictions to the ground truth. We select a band of $1.0 \mathrm{"/yr}$ about $\dot \omega_6$ and $\dot \Omega_6$ and consider all the asteroids in this band as being in the secular resonance.  The average of the observed albedos in this region is $p_V = 0.07$ whereas the average predicted albedo is slightly larger at $p_V = 0.08$.  This suggests our model generalizes rather well even to situations where it has not been trained. However, as can be readily seen in Figure \ref{familyvis}, contamination from nearby sources is significant and hence the diversity in the albedo distribution in secular resonances may be overestimated due to this contamination. 

\begin{figure}[!ht]
\centering
    \includegraphics[totalheight=7.5cm]{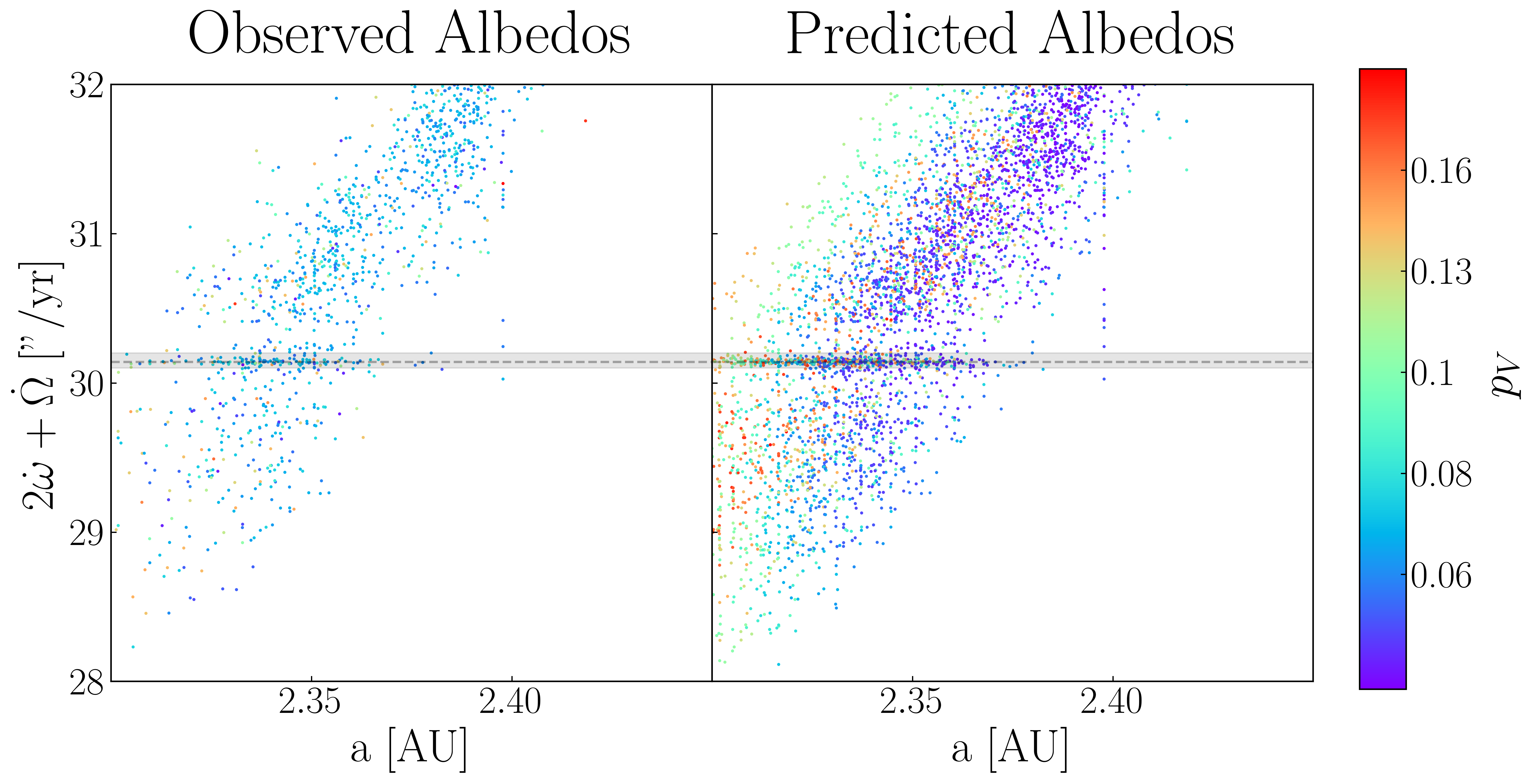}
    \caption{
    We show the observed $\log_{10}(p_V)$ albedos of asteroids near the z2 resonance in the left panel, and the predicted $\log_{10}(p_V)$ albedos of asteroids in the same region in the right panel, shown in the plane of proper $a$ and $2 \dot \omega + \dot \Omega$ proper precessional frequencies.  We outline the secular resonance with a horizontal dashed line $2 \dot \omega_6 + \dot \Omega_6$ and consider all those asteroids in the shaded region as part of the resonance.}
    \label{erigone}
\end{figure}

\section{Conclusion}

    In this paper, we developed a neural network based model for predicting asteroid albedos from their proper elements. Our model predicts asteroid albedos significantly more accurately than the naive approach of assuming an average albedo for the entire belt, lowering the mean absolute error by around 37\% compared to that approach.  We find our model's predictive power isn't limited to asteroid families with homogeneous albedos but also fits weaker trends over the more diverse background asteroids in the main belt. As a consequence of this fitting, the model also uncovers a large amount of structure in the asteroid belt, with many of these structures correlating with known asteroid families, with the relationship between these regions and the asteroid families a subject for future work. 
    This modeling may prove invaluable as future surveys discover more asteroids for which albedo measurements are not otherwise available.

\section{Acknowledgements}

We are grateful to Matt Holman for helpful discussions.  This publication makes use of data products from the Near-Earth Object Wide-field Infrared Survey Explorer (NEOWISE), which is a joint project of the Jet Propulsion Laboratory/California Institute of Technology and the University of Arizona. NEOWISE is funded by the National Aeronautics and Space Administration. Finally, we are thankful for the input of the anonymous reviewers, whose feedback and suggestions significantly improved this manuscript.  

\section{Data Availability}

In addition to our neural net ensemble, we also produce a catalog created by using our trained model on the full set of asteroids with proper elements given by the Asteroid Families Portal. The Asteroid Families Portal contains 585,174 unique numbered main belt asteroids with recorded proper elements and predicted albedos.  Hence this catalog extends the NEOWISE albedos by nearly a factor of five. This catalog includes entries for the mean neural net prediction, $\sigma_{belt}$ and $\sigma_{pred}$ for both the visible and infrared albedos and is available doi:\dataset[10.5281/zenodo.7796841]{10.5281/zenodo.7796841}.  The weights used by the neural net are available on GitHub via \url{https://github.com/r-zachary-murray/Asteroid-Albedos}.

\bibliography{refs}{}

\bibliographystyle{aasjournal}

\end{document}